\documentclass[aps, twocolumn, tightenlines, ece, amssymb, amsfonts, amsmath, groupedaddress]{revtex4}
\usepackage{epsf}
\usepackage{graphics}
\usepackage{graphicx}
\usepackage{epstopdf}

\begin{document}

\title{Griffiths phase and critical behavior in layered Sr$_2$IrO$_4$ ferromagnet}

\author{A. Rathi$^{a,b}$, Sonam Perween$^{a,b}$, P. K. Rout$^{a}$, R. P. Singh$^{c}$, Anurag Gupta$^{a,b}$,  Sukhvir Singh$^{a,b}$,
B. Gahtori$^{a,b}$, B. Sivaiah$^{a,b}$, Ajay Dhar$^{a,b}$, A. K.
Shukla$^{a,b}$, R. K. Rakshit$^{a,b}$, R. P. Pant$^{a,b}$ and G. A.
Basheed$^{a,b,\dag}$}

\affiliation{$^{a}${CSIR-National Physical Laboratory (NPL), Dr. K.
S. Krishnan Marg, New Delhi - 110012, India}\\
$^{b}${Academy of Scientific and Innovative  Research (AcSIR), NPL
Campus, New Delhi - 110012, India}\\
$^{c}${Department of Physics, Indian Institute of Science Education
and Research (IISER) Bhopal, Bhopal-462 023, India}}
\email[$\dag$]{basheedga@nplindia.org}

\begin{abstract}
We report the existence of  Griffiths phase (GP) and its influence
on critical phenomena in layered Sr$_2$IrO$_4$ ferromagnet (T$_C$ =
221.5 K). The power law behavior of inverse magentic susceptibility,
1/$\chi$(T) with exponent $\lambda = 0.18(2)$ confirm the GP in the
regime T$_C$ $<$ T $\leq$ T$_G$ = 279.0(5) K. Moreover, the detailed
critical analysis via modified Arrott plot method exhibits
unrealistic critical exponents $\beta$ = 0.77(1), $\gamma$ = 1.59(2)
and $\delta = 3.06(4)$, in corroboration with magneto-caloric study.
The abnormal exponent values have been viewed in context of {\it
ferromagnetic-Griffiths} phase transition. The GP has been further
analyzed using Bray model, which yields a reliable value of $\beta$
= 0.19(2), belonging to the two-dimensional (2D) \emph{XYh$_4$}
universality class with strong anisotropy present in Sr$_2$IrO$_4$.
The present study proposes Bray model as a possible tool to
investigate the critical behavior for Griffiths ferromagnets in
place of conventional Arrott plot analysis. The possible origins of
GP and its correlation with insulating nature of Sr$_2$IrO$_4$ have
been discussed.

\vspace{1pc}\noindent{\it Keywords}: {\it 5d} iridates, Griffiths
phase, Critical phenomena.

\end{abstract}

\maketitle

\section{INTRODUCTION}

The Ir-based {\it 5d} transition metal oxides (TMOs) are widely
known to exhibit various unconventional and novel properties
{\cite{Kim1, Wang2, Wan3}}, driven by strong spin-orbit coupling
(SOC). The spatially extended nature of {\it 5d} electrons in these
oxides results in weak (compared to {\it 3d} TMOs) electronic
correlations and a metallic paramagnetic (PM) behavior is
anticipated. In contrast, many iridates {\cite{Cao4, Crawford5}}
happen to be magnetic insulators due to large SOC present in the
system. In particular, the layered iridate Sr$_2$IrO$_4$ (SIO) has
emerged as an interesting system due to its structural
{\cite{Crawford5}} and magnetic {\cite{Kim1}} similarities with
the novel layered cuprate spin-1/2 Heisenberg antiferromagnet (AFM)
La$_2$CuO$_4$. The large SOC ($\sim$ 0.4 eV) in SIO splits the
Ir$^{4+}$ (d$^5$) t$_{2g}$ level into Kramer's doublet with a filled
J$_{eff}$ = 3/ 2 low energy state and a half filled effective
J$_{eff}$ = 1/2 high energy state {\cite{Kim6}}, which makes the
system different from the spin-1/2 La$_2$CuO$_4$.

The SIO crystallizes in  a tetragonal crystal structure with an
arrangement of two-dimensional (2D) IrO$_2$ layers along the c-axis,
as in La$_2$CuO$_4$, but with the rotation of IrO$_{6}$ octahedra
around c axis by $\sim$ 11$^0$ {\cite{Crawford5, Huang7}}. This
rotation breaks the inversion symmetry of in-plane Ir-O-Ir network
and Dzyaloshinsky-Moriya (DM) interaction between J$_{eff}$ = 1/2
spins gives rise to a weak ferromagnetic (FM) order below T$_C$
$\sim$ 230 K with a moment of $\sim$ 0.14 $\mu$$_B$/Ir within each
IrO$_2$ (basal ab-) plane {\cite{Crawford5, Ye8}}. The system
otherwise would behave like a collinear AFM, as La$_2$CuO$_4$.
Furthermore, the X-ray resonant magnetic scattering (XRMS) studies
on SIO established the correlation between J$_{eff}$ spins in PM
state {\cite{Fujiyama9, Vale10}}. This often leads to ``{\it
Griffiths singularity}" {\cite{Griffiths11, Salamon12}}, which is
characterized by the presence of short-range FM clusters above Curie
temperature.

Griffiths  {\cite{Griffiths11}}, originally, proposed the
``{\it Griffiths singularity}" to explain the role of quenched
randomness in diluted Ising ferromagnets. Bray et al. {\cite{Bray13,
Bray14}} generalized the argument for any disordered system, where
intrinsic disorder breaks the system into short-range FM
clusters of different sizes with each one having its own Curie
temperature, T$_C$ (p), where p is the degree of disorder. The
long-range FM order is established only below a threshold disorder,
p$_C$ at a much lower temperature, T$_C^R$ compared to the zero
disorder system Curie temperature, T$_G$, the Griffiths temperature.
The temperature regime, T$_C^R$ $<$ T$_C$ (p) $\leq$ T$_G$ (p = 0),
in which the singularity exists, is usually referred to the
Griffiths phase (GP).

Here, for the first time, we report Griffiths phase in layered
Sr$_2$IrO$_4$ ferromagnet. The essential requirement for realization
of GP is the absence of long-range order, which can be confirmed
through critical phenomena analysis. In this paper, we present the
critical behavior across the {\it ferromagnetic-Griffiths} phase
transition in Sr$_2$IrO$_4$ via {\it dc} magnetization measurements.
The basic difference between pure (conventional) and Griffiths
ferromagnets lies in their free energy expression across the phase
transition; the former has an analytic free energy expression while
the expression for later is non-analytic {\cite{Griffiths11}}. The
non-analytic nature of GP leads to unrealistically large critical
exponents {\cite{Jiang15, Phan16, Triki17}} through modified {\it
Arrott} plot, scaling law and magneto-caloric analysis. Here, we
utilize Bray model to investigate the critical behavior of
Sr$_2$IrO$_4$. The present study proposes Bray model as a possible
tool to investigate the critical behavior for Griffiths ferromagnets.
We also discuss the possible origin of insulating nature of Sr$_2$IrO$_4$ in
the context of GP.

\begin{figure}
\begin{center}
\vspace{-0.5em}
\includegraphics[scale=0.34]{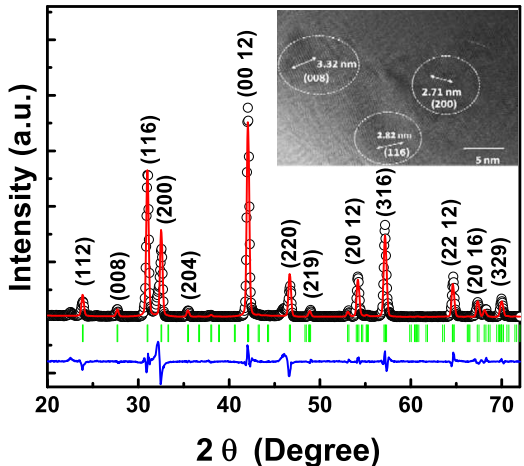}
\vspace{-1.5em}
\caption{Room temperature X-ray diffraction pattern
(open circles) of Sr$_2$IrO$_4$, along with the full-profile
(Rietveld) fit (solid red line). The difference between observed and
calculated data is shown by solid blue line. The vertical bars show
the position of allowed (hkl) Bragg reflections. The inset shows the
HRTEM image of Sr$_2$IrO$_4$.}
\vspace{-2.0em}
\end{center}
\end{figure}

\begin{figure}
\begin{center}
\includegraphics[scale=0.45]{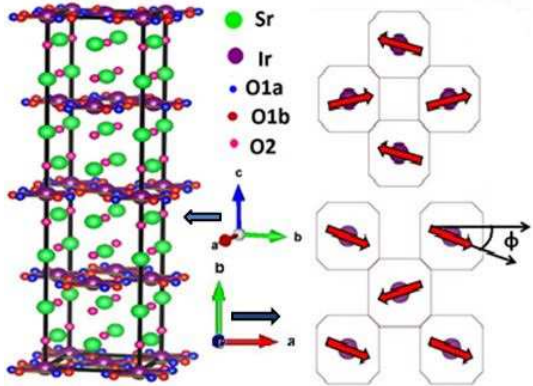}
\vspace{-1.em}
\caption{The left panel shows the layered pervoskite
structure of Sr$_2$IrO$_4$. The right panel shows the schematic
magnetic structure of Sr$_2$IrO$_4$ with the clockwise and
anticlockwise rotation of alternate basal (ab) planes around c-axis
by $\phi$ $\sim$ 11$^0$.}
\vspace{-3.0em}
\end{center}
\end{figure}

\section{Experimental Details}
\vspace{-1.0em}

The Sr$_2$IrO$_4$ polycrystalline sample was
synthesized through solid state reaction method. The starting oxides
SrCO$_3$ (99.99\% purity) and IrO$_2$ (99.99\% purity) were mixed in
stoichiometric ratio. The mixture was heated in air at 800, 900 and
1000$^{0}$C for 12 hours each with intermediate grindings. Finally,
the mixture was pressed into pellet and sintered at 1000$^{0}$C for
24 hours. The phase quality of the sample was examined using a
four-circle high resolution X-ray diffractometer (Cu-K$_\alpha$
radiation). Figure 1 shows the room temperature X-ray diffraction
pattern of Sr$_2$IrO$_4$, along with the profile (Rietveld) fit
using FULLPROF program. The Rietveld fit confirms the single
tetragonal phase (space group I4$_1$/acd) with the refined lattice
parameters, a = 5.50 {\AA}, b = 5.50 {\AA} and c = 25.76 {\AA},
which are in close agreement with earlier reports on SIO
{\cite{Crawford5, Huang7}}. The inset of Fig. 1 shows the HRTEM
image of as synthesized SIO sample which clearly reveals (008),
(116) and (200) atomic planes which further confirms high
crystallinity of Sr$_2$IrO$_4$ sample. The Rietveld fit was further
used to generate the crystal structure and (schematic) magnetic
structure. The left panel of Fig. 2 shows the layered pervoskite
structure of Sr$_2$IrO$_4$, while the right panel reveals the
magnetic structure with rotation of $\phi$ $\sim$ 11$^0$ of
alternate basal (ab-) planes around c axis, in  clockwise and
anticlockwise direction, respectively.  The magnetic properties were
investigated by measuring the detailed dc magnetization, M (H, T),
using SQUID magnetometer. Furthermore, the magnetic isotherms were
recorded in the vicinity of critical temperature using a Quantum
Design PPMS.

\begin{figure}
\begin{center}
\vspace{-2.75em}
\includegraphics[scale=0.33]{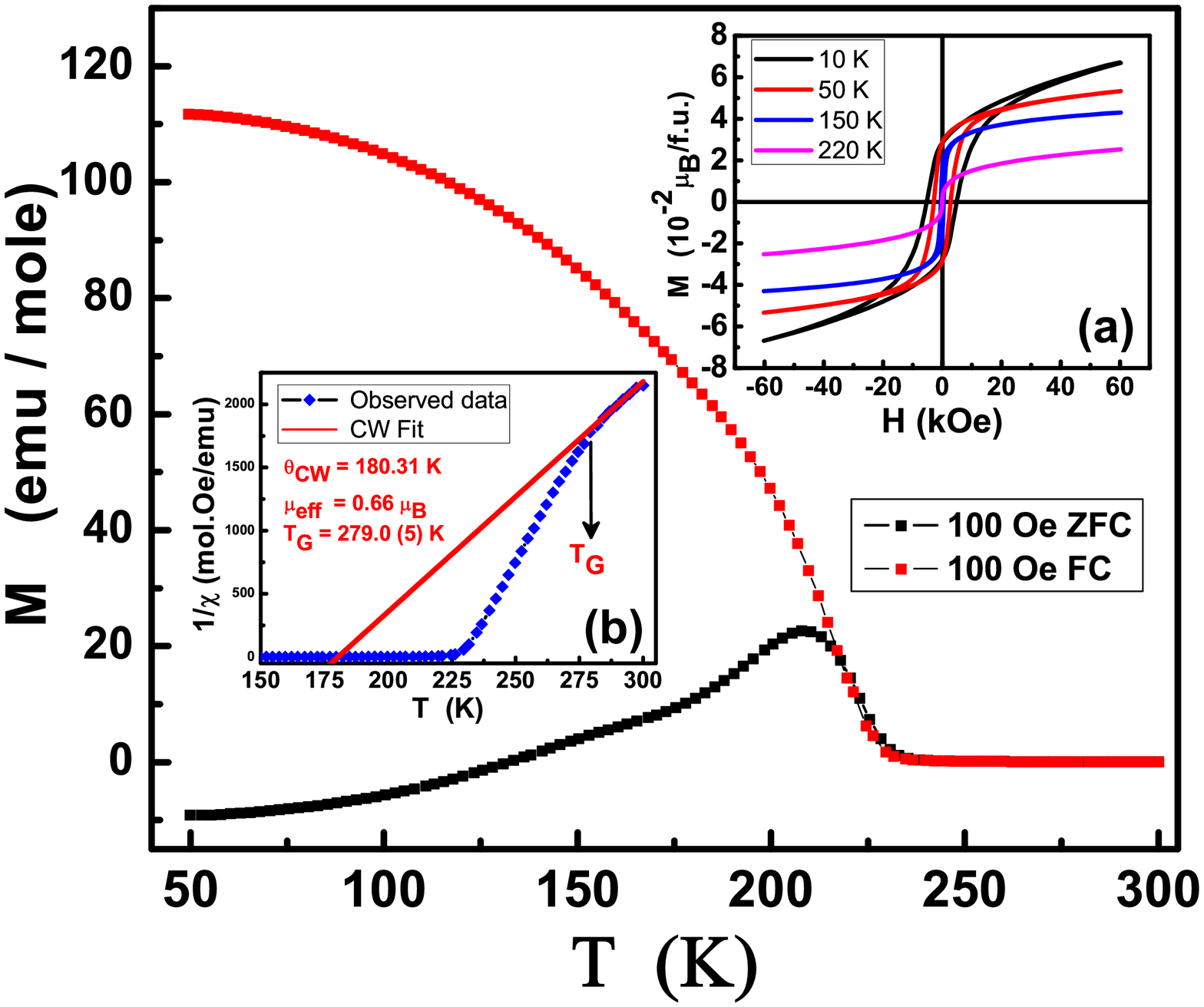}
\vspace{-3.0em}
\caption{The temperature dependence of dc
magnetization in FC and ZFC modes at 100 Oe. Inset (a) shows the MH
loops observed at different temperatures. Inset (b) shows 1/ $\chi$
(T) plot along with Curie-Weiss law fit.}
\vspace{-3.0em}
\end{center}
\end{figure}

\section{Result and discussion }
\subsection{dc magnetization}

Figure 3 shows the temperature dependence of dc magnetization
measured in the field-cooled (FC) and zero-field-cooled (ZFC) modes
at 100 Oe. The plot shows the sharp increase in magnetization below
T$_C$ $\sim$ 220 K, which is in close agreement with earlier reports
{\cite{Crawford5, Kini18, Bhatti19}}. On cooling, the bifurcation
between FC and ZFC magnetization appears, along with a broad maxima
in ZFC magnetization, which suggests a large magnetic anisotropy in
the system. On further cooling, the dc magnetization shows minor AFM
transition at $\sim$ 160 K, which is consistent with earlier reports
on SIO {\cite{Bhatti19, Chikara20}}. Moreover, the AFM transition
weakens with the increasing field and suppresses in a high field of
1 T (not  shown here), which can be associated with domination of FM
moment over AFM moment in high fields. The inset (a) in Fig. 3 shows
the MH loops measured at different temperatures with a field range
of $\pm$ 6 T. The loop at 10 K shows the non-saturating behavior
with a magnetization value of $\mu$$_H$ = 0.07 $\mu$$_B$/f.u. at 6
T. The observed $\mu$$_H$ value is consistent with earlier reports
{\cite{Bhatti19}}, however, much smaller compared to the theoretical
value, $\mu$$_H$ = g J$_{eff}$ $\mu$$_B$ = 0.33 $\mu$$_B$/f.u. with
J$_{eff}$ = 1/2 and g$_J$ = 2/3. The non-saturating behaviour and
very small $\mu$$_H$ value confirm the presence of low temperature
AFM component. Moreover, the observed high remenance (M$_R$ $\sim$
0.028 $\mu$$_B$/f.u.) and coercivity (H$_C$ $\sim$ 5.20 kOe) at 10 K
confirm large magnetic anisotropy present in the system. The MH
loops observed at higher temperatures, from 50 K to 220 K show that
the saturating behavior increases with increase in temperature due
to the lowering of AFM contribution.

The inset (b) in Fig. 3 shows the temperature dependence of inverse
magnetic susceptibility (1/$\chi$) at 100 Oe. The high temperature
data, fitted with the Curie-Weiss (CW) equation, $\chi = C / (T -
\theta_{CW})$, gives $\theta$$_{CW}$ = +180.31 K and $\mu$$_{eff}$ =
0.66 $\mu$$_B$. The positive $\theta$$_{CW}$ value confirms the FM
ordering, whereas the obtained $\mu$$_{eff}$ value exceeds the
theoretical value $\mu_{eff} = g_J \sqrt{J_{eff}(J_{eff} + 1)}$ =
0.57 $\mu$$_B$/f.u. More importantly, the 1/$\chi$ (T) plot shows
the negative deviation from the CW  behavior much above T$_C$ $\sim$
220 K. The negative downturn in 1/$\chi$ plot above T$_C$ and the
larger experimental $\mu$$_{eff}$ value suggest the formation of
short range magnetic clusters in PM state, which is a signature of
Griffiths singularity {\cite{Salamon12, Jiang21}}. The Griffiths
temperature can be determined as T$_G$ = 279.0(5) K and the
temperature range of GP with respect to pure FM phase is calculated
as GP \% = (T$_G$-T$_C$)/T$_C$ $\times$ 100 which comes out to be
26.8 \%. The GP has been discussed in more detail in section III (C).

\begin{figure}
\begin{center}
\vspace{-2.0em}
\includegraphics[scale=0.35]{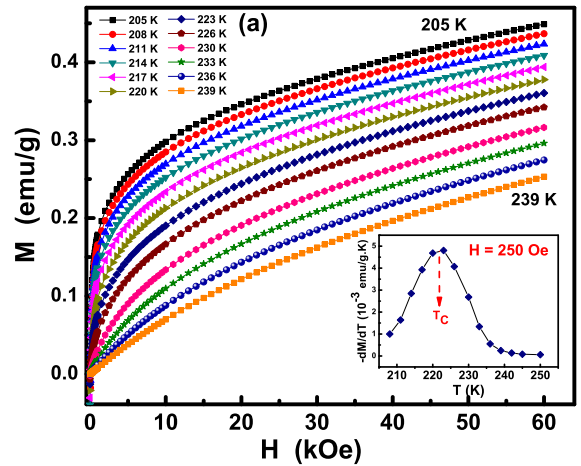}
\includegraphics[scale=0.35]{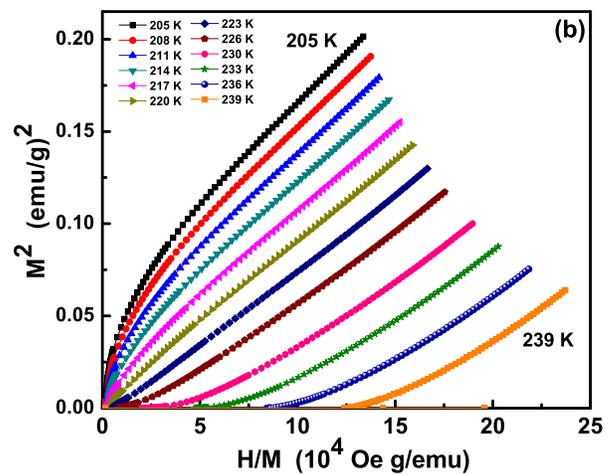}
\vspace{-3.0em}
\caption{(a) The magnetic isotherms at different
temperatures around T$_C$. The inset shows the - dM/dT versus T plot
at 250 Oe, showing magnetic transition at 220 K $\leq$ T$_C$ $\leq$ 223 K. (b) The basic Arrott plot; M$^2$ against H/M
(Mean field model, $\beta$ = 0.5 and $\gamma$ = 1.0) at different temperatures around T$_C$.}
\vspace{-3.0em}
\end{center}
\end{figure}

\subsection{Critical phenomena}

The basic feature of GP is the absence of long-range magnetic order
in the system, which can be examined through Arrott plot analysis.
Figure 4(a) shows the M (H) isotherms at different temperatures
around T$_C$, ranging from 205 K to 239 K with a temperature
interval of 3 K. All the isotherms show non-linear and
non-saturating behavior up to 6 T, indicating the presence of
short-range FM order {\cite{Ho22}}. The inset in Fig. 4(a) shows the
- dM/dT versus temperature plot at 250 Oe, revealing the T$_C$
value, lying between 220 and 223 K. Figure 4(b) shows the basic
Arrott plots {\cite{Arrott23}}; M$^2$ against H/M (Mean filed model
{\cite{Kaul24}}) for the temperatures near T$_C$. The Arrott plots
have positive slope, which confirms the second-order magnetic phase
transition (SOPT), according to Banerjee's criteria
{\cite{Banerjee25}}. Furthermore, a series of parallel and straight
lines are expected for high field data in Arrott plot, however, the
plots show a non-linear diverging behaviour with increasing magnetic
field. The breakdown of long-range mean field model supports the
presence of Griffiths phase. Therefore, the modified Arrott plot
(MAP) method has been used, which is based on the Arrott-Noakes
equation, $(H/M)^{1/\gamma} = (T-T_C)/T_1 + (M/M_1)^{1/\beta}$,
where $\beta$ and $\gamma$ are the critical exponents; T$_1$ and
M$_1$ are constants {\cite{Noakes26}}. The two critical exponents
are further related through Widom scaling relation, $\delta = 1 +
\gamma/\beta$ {\cite{Widom27}} where $\delta$ is third critical
exponent. The modified Arrott plots; M$^{1/\beta}$ against
(H/M)$^{1/\gamma}$ were constructed using critical exponents for
different known universality classes (Table I). However, the plots
(not shown here) do not show the expected parallel and straight
lines, excluding their validity for SIO system and hence, a detailed
critical analysis is required.

For a SOPT, the critical exponents ($\beta$, $\gamma$ and $\delta$)
can be determined from the spontaneous magnetization, M$_S$ (T $<$
T$_C$), initial susceptibility, $\chi$$_0$ (T $>$ T$_C$) and the
critical isotherm, M (H, T$_C$) asymptotic relations
{\cite{Noakes26}},

\begin{equation}
M_S(T) = M_0 (-\varepsilon)^\beta,~~~~~~T < T_C
\end{equation}

\begin{equation}
1/\chi_0 = (h_0/M_0)(\varepsilon)^\gamma, ~~~~~~T > T_C
\end{equation}

\begin{equation}
M = D H^{1/\delta}, ~~~~~~T = T_C
\end{equation}

\begin{figure}
\begin{center}
\includegraphics[scale=0.31]{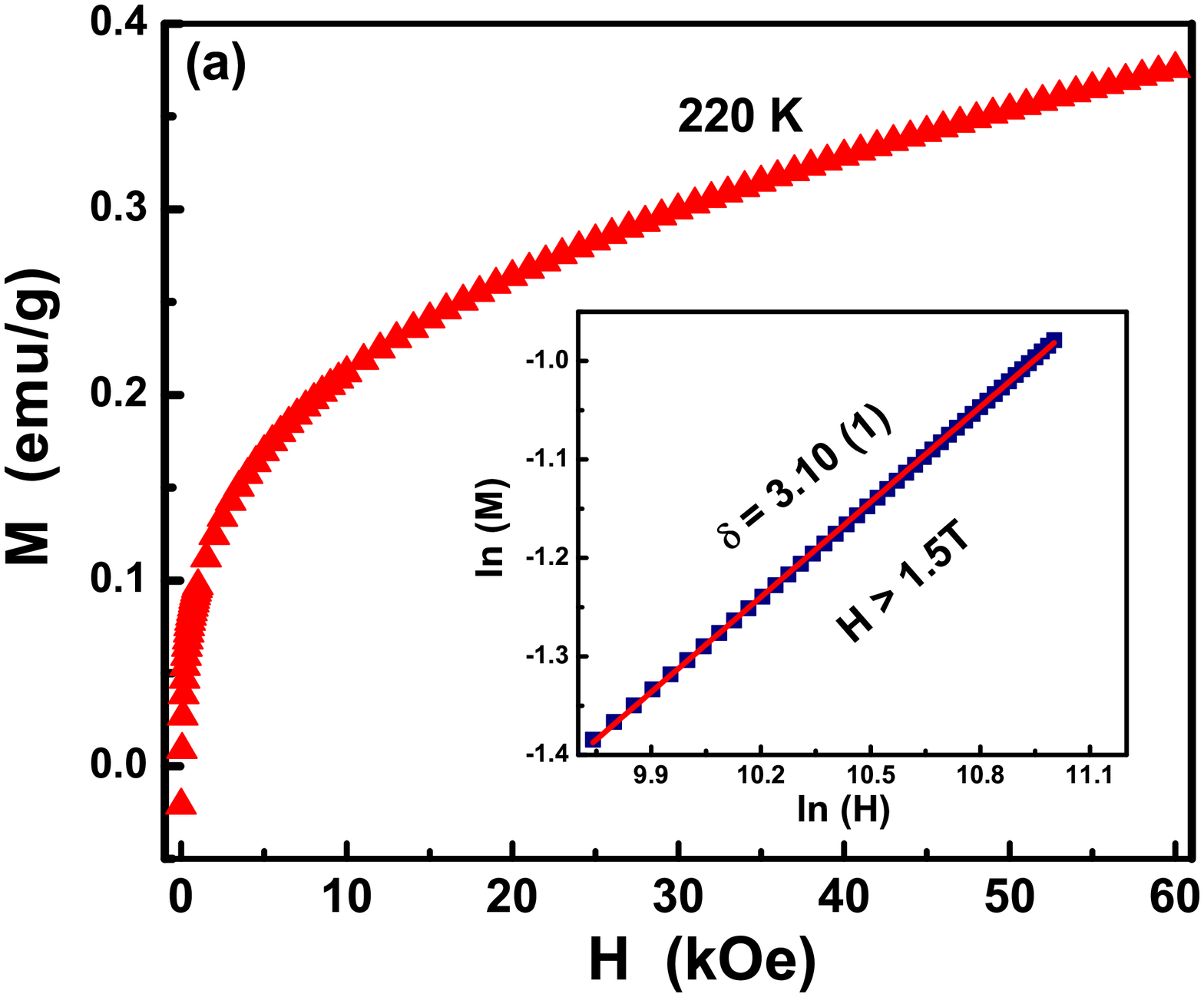}
\includegraphics[scale=0.31]{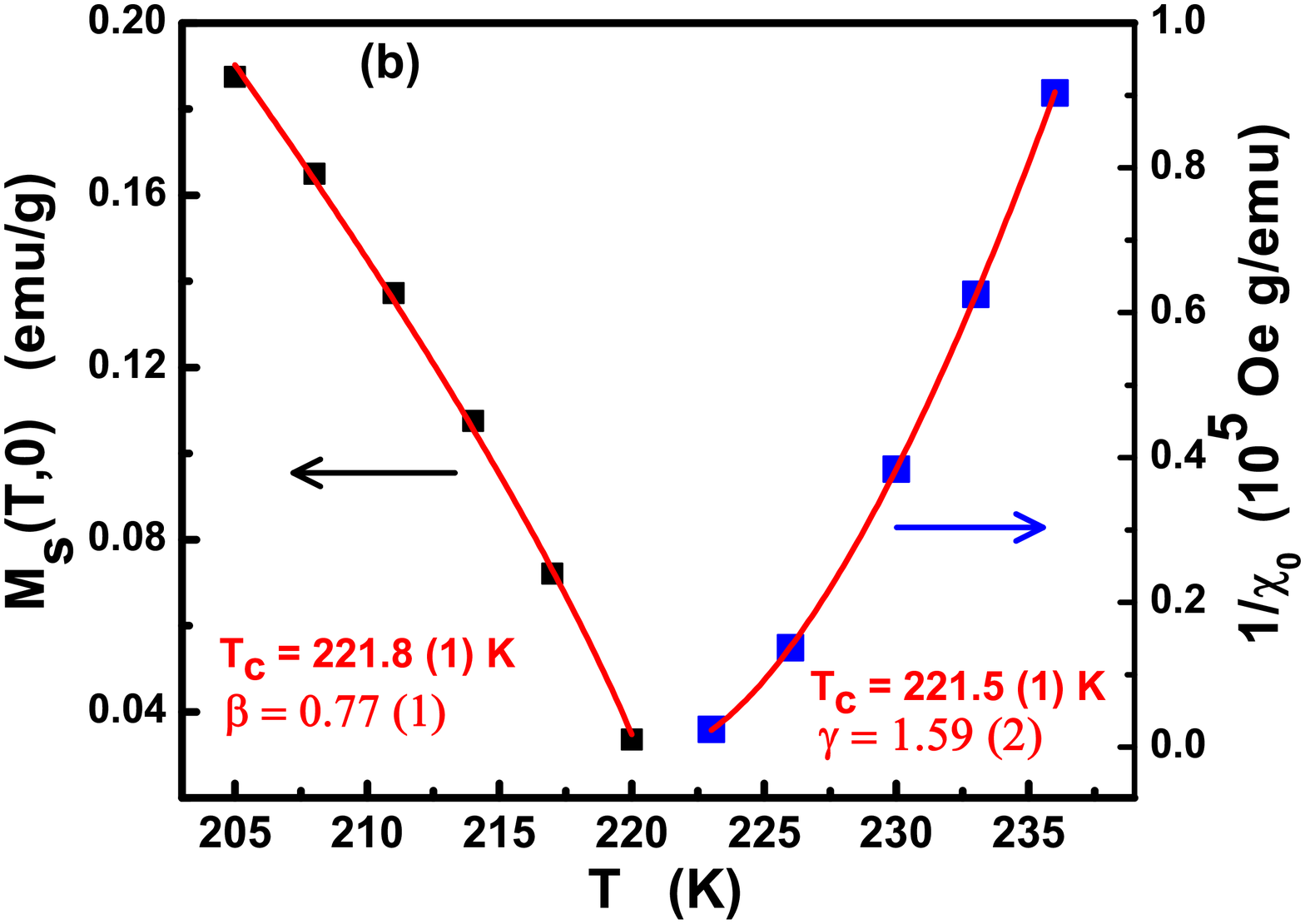}
\vspace{-2.0em}
\caption{(a) Magnetic isotherm M(H) at T = 220 K near
T$_C$ to calculate the critical exponent $\delta$ value. The inset
shows the same plot on log-log scale with linear fit (solid red line). (b) The temperature dependence of spontaneous
magnetization, M$_S$ (T, 0) and the inverse initial susceptibility
1/$\chi$$_0$ (T), along with power law (Eq. 1 and 2) fits (solid curves).}
\vspace{-3.0em}
\end{center}
\end{figure}

where $\varepsilon = (T - T_C)/T_C$ is the reduced temperature;
M$_0$, h$_0$ and D are the critical amplitudes. The reliable
critical exponents have been determined using an iterative
self-consistent method. Firstly, the $\delta$ value has been
determined using critical isotherm (CI) method (Eq. 3) by plotting
the d ln(H)/d ln(M) against H for all the isotherms. The isotherm
with its slope closest to zero corresponds to T = T$_C$ and the
intercept gives the $\delta$ value. Accordingly, T$_C$ = 220(3) K
and $\delta$ = 3.1(2) have been obtained. Figure 5(a) shows the
magnetic isotherm at T = 220 K (near T$_C$), along with the same plot on log-log
scale in the inset. Furthermore, the exact critical exponents follow
the scaling law {\cite{Noakes26}} in the critical region, according
to which, the M(H, $\varepsilon$)/$\mid\varepsilon\mid^{\beta}$
against H/$\mid\varepsilon\mid^{\beta\delta}$ plots lie on two
universal branches below and above T$_C$. Using this approach, the
$\beta$ and $\delta$ values are obtained as 0.76(3) and 3.0(1),
respectively, using T$_C$ = 221 K. The $\gamma$ value was determined
as 1.52(8) using the Widom scaling relation {\cite{Widom27}}. The
obtained $\beta$ and $\gamma$ values were used as initial trial
exponents for the MAP analysis. The linear extrapolation of the high field MAP data
gives M$_S$(T) and 1/$\chi$$_0$ (T) values as intercept on M$^{1/\beta}$ and
(H/M)$^{1/\gamma}$ axes, respectively. The M$_S$(T) and 1/$\chi$$_0$ (T) values were used for
fitting of Eq. 1 and Eq. 2, respectively. The best fits to the data
give $\beta$ = 0.77(1) with T$_C$ = 221.8(1) and $\gamma$ = 1.59(2) with T$_C$ = 221.5(1),
respectively [See Fig. 5(b)]. Moreover,
the $\delta$ value can be determined as 3.06(4) using the Widom
scaling relation.

\begin{figure}
\begin{center}
\includegraphics[scale=0.35]{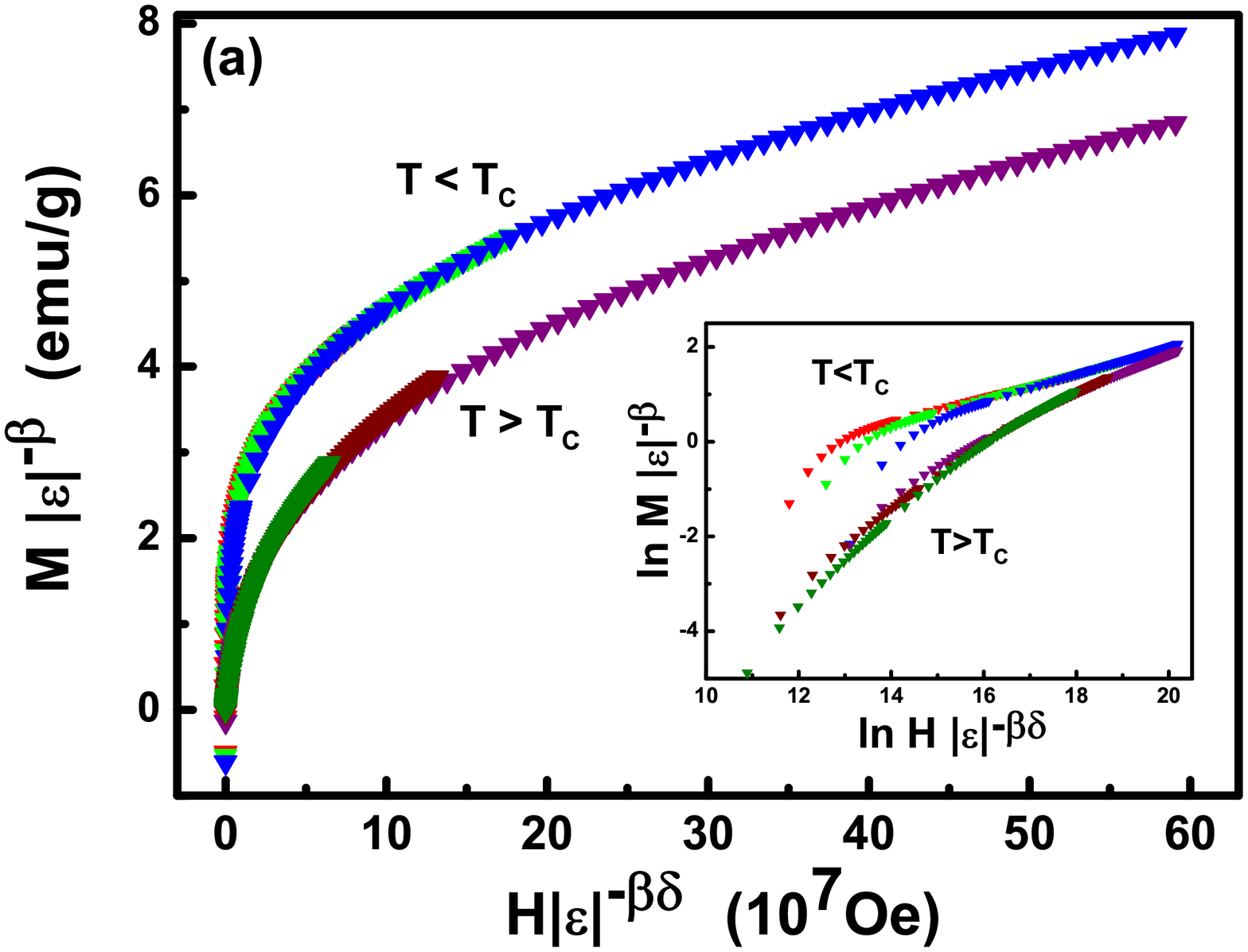}
\includegraphics[scale=0.35]{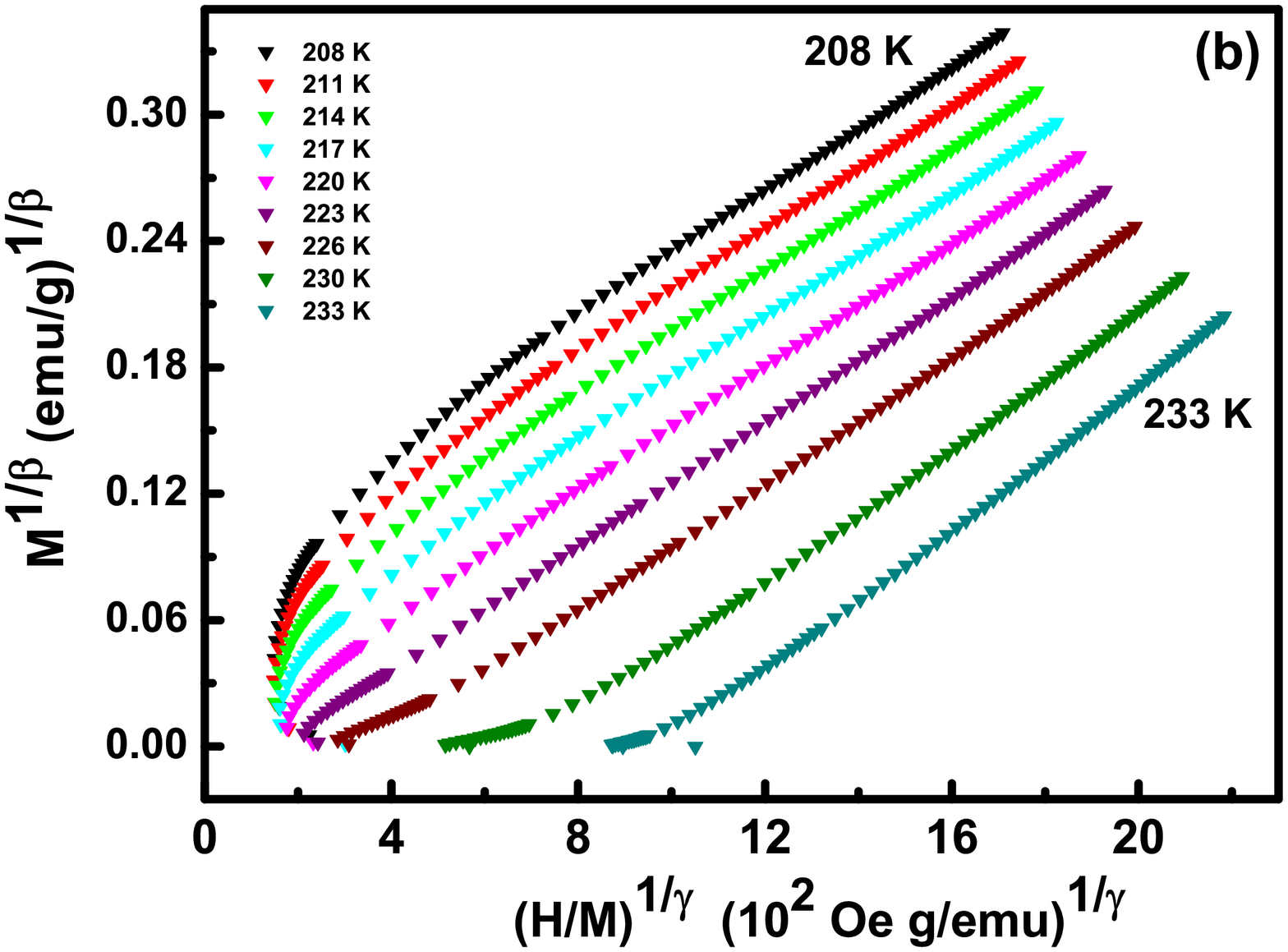}
\vspace{-3.0em}
\caption{(a) The scaling plots, below and above T$_C$
using exponents from the modified Arrott plot analysis, along with
same plot on log-log scale in the inset. (b) The Modified Arrott plot; M$^{1/\beta}$
against (H/M)$^{1/\gamma}$ with $\beta$ = 0.77 and $\gamma$ = 1.59,
at different temperatures around T$_C$.}
\vspace{-2.0em}
\end{center}
\end{figure}

The critical exponents from MAP method were examined using scaling
theory with T$_C$ = 221.5 K. Figure 6(a) shows two universal
branches for T $>$ T$_C$ and T $<$ T$_C$ respectively, which
reflects the reliability of critical exponents. The modified Arrott
plot reconstructed using above critical exponents is shown in Fig.
6(b). The MAP shows nearly parallel straight lines for high field
region (H $\geq$ 0.5 T) in the vicinity of T$_C$. The low field
non-linear behavior can be explained as a result of contribution
from different magnetic domains, aligned along different directions
{\cite{Pramanik28}}.

\begin{figure}
\begin{center}
\includegraphics[scale=0.31]{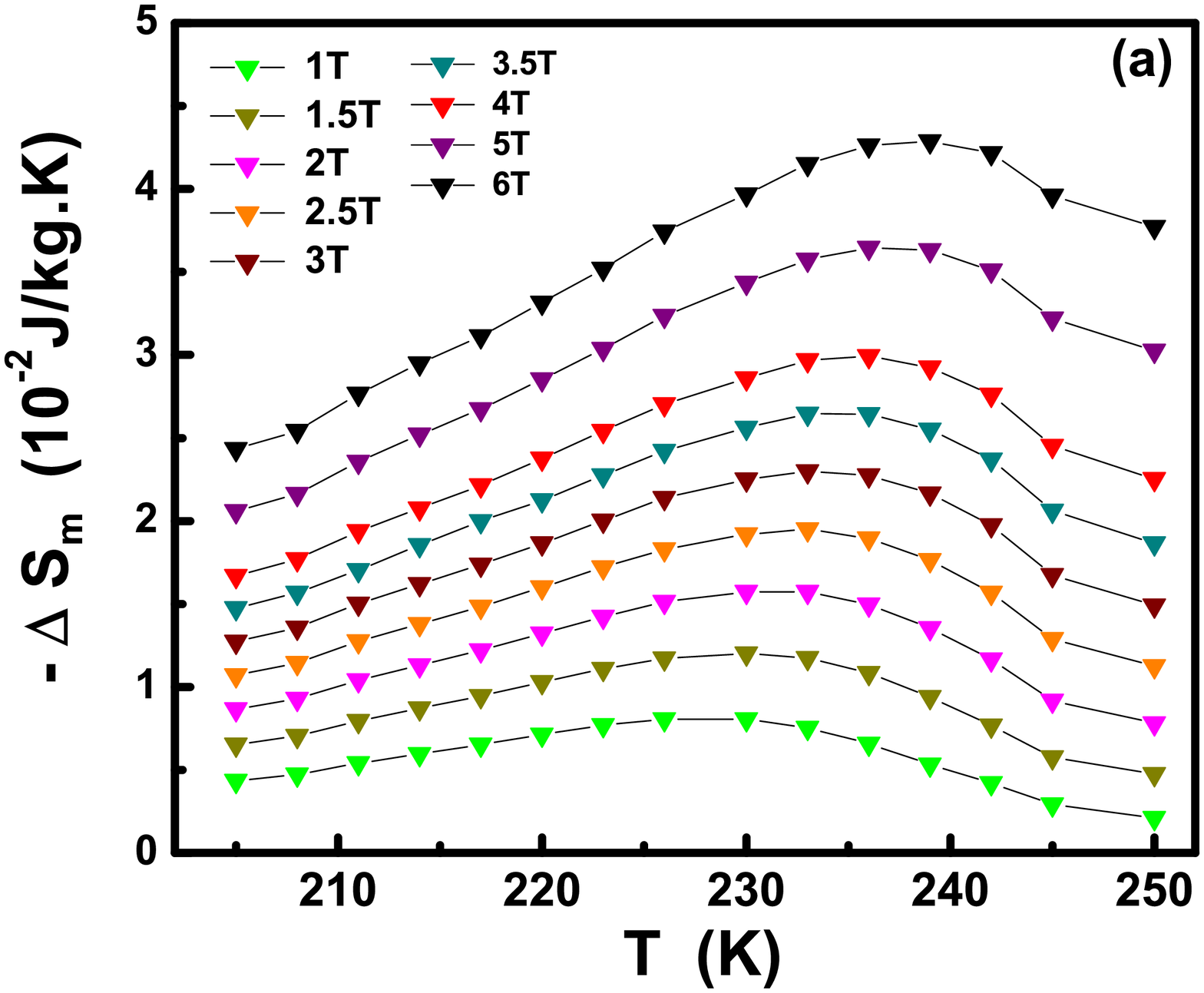}
\includegraphics[scale=0.33]{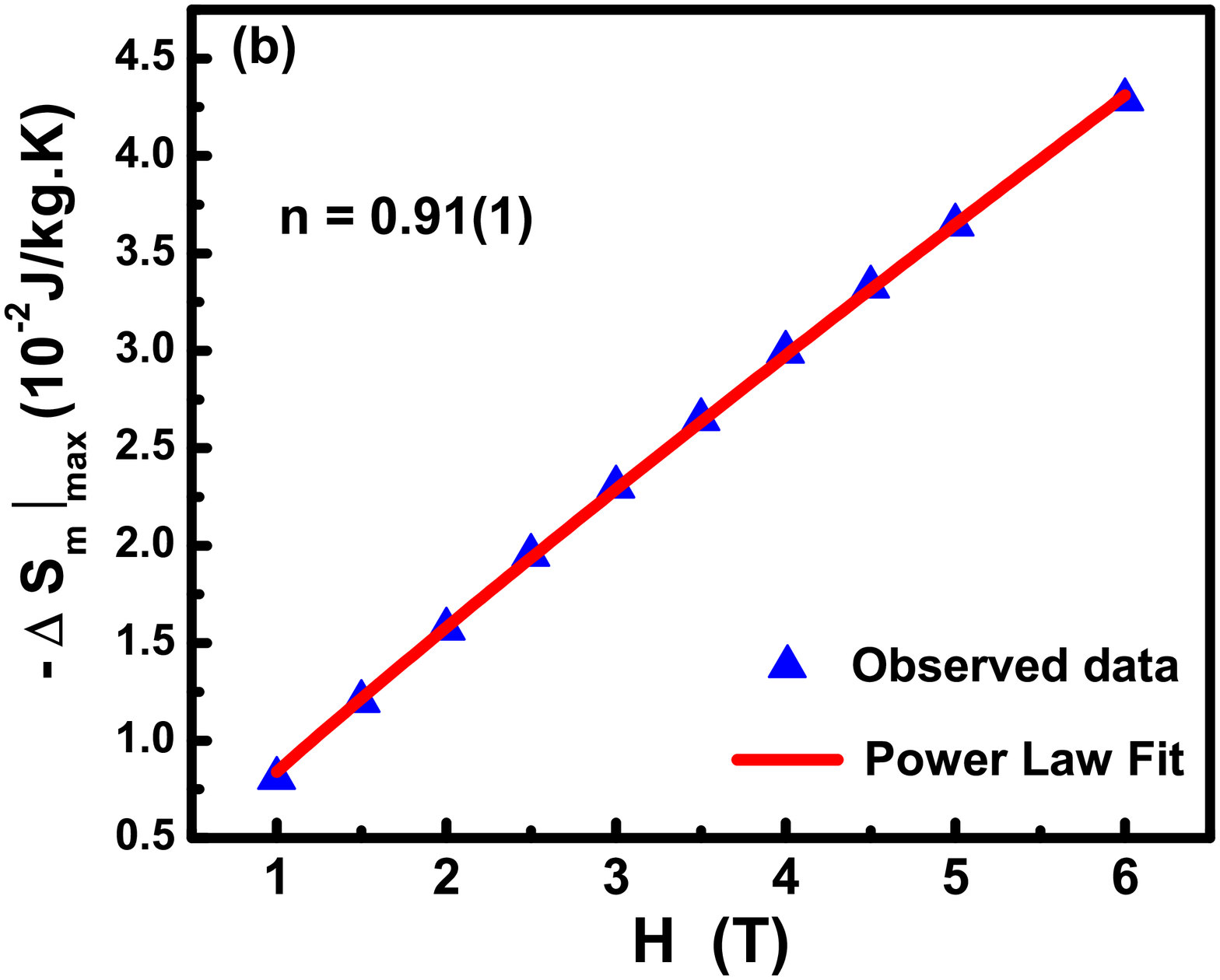}
\vspace{-2.0em}
\caption{(a) The change in magnetic entropy, -
$\Delta$S$_m$ (T) for different applied magnetic fields. (b) The magnetic field dependence of - $\Delta$
S$_m$$\mid$$_{max}$, along with power law [$ - \Delta S_m\mid_{max} \propto H^n$] fit.}
\vspace{-2.0em}
\end{center}
\end{figure}

The obtained critical exponents were further examined by analyzing
the magneto-caloric effect in SIO across the phase transition. The
change in magnetic-entropy ($\Delta$S$_m$) is calculated using
Maxwell relation {\cite{Franco29}} as

\begin{equation}
\Delta S_m = \left(\frac{\partial M}{\partial T}\right)_{H} \times H
\end{equation}

Figure 7(a) shows the temperature dependence of the calculated -
$\Delta$S$_m$ for different fields. As, the - $\Delta$S$_m$ is
related to the change in magnetization with temperature, which is
sharpest at the critical point, the - $\Delta$S$_m$ plot shows broad
peak around the critical temperature. The peak values of -
$\Delta$S$_m$ at the critical temperature for different fields are
shown in Fig. 7(b). This data have been analyzed through power law,
$ - \Delta S_m\mid_{max} \propto H^n$, where n is the magnetic-ordering parameter.
The best fit gives 'n' value as 0.91(1), which
is consistent with n = 0.90, calculated using the theoretical
relation, $n = 1 + (\beta-1)/(\beta+\gamma)$ {\cite{Ho22,
Franco29}}. Thus, the magneto-caloric study further confirms the
reliability of obtained critical exponent values.

\begin{table}[t]

 \caption{Comparison of critical exponents of Sr$_2$IrO$_4$ from MAP, CI and GP analysis with reported values and different theoretical models.}
\begin{ruledtabular}
\begin{tabular}{ccccccc}
 Material/Model& Ref   & Method       & $\beta$    & $\gamma$ & $\delta$\\
\\
\hline
SIO & This Work & MAP   & 0.77(1) & 1.59(2)& 3.06(4)\\
SIO &This Work & CI & - & - &3.1(2)\\
SIO &This Work & GP & 0.19(2) & - &-\\
SIO &{\cite{Miyazaki30}} & $\mu$SR & 0.20 & - &-\\
SIO &{\cite{Ye8, Chetan33}} & ND & 0.18 & - &-\\
SIO &{\cite{Vale10}} & XRMS & 0.195 & - &-\\
{d:n}={3:3}\\
Mean Field  & {\cite{Kaul24}} & Theory  &0.5&1.0&3.0\\
3D Heisenberg  & {\cite{Kaul24}} & Theory & 0.365 & 1.386 &4.8 \\
{d:n}={3:1}\\
3D Ising  & {\cite{Kaul24}}& Theory  & 0.325 & 1.24&4.82\\
{d:n}={2:2}\\
2D XY & {\cite{Nielsen31}} & Theory  & 0.23 & -&-\\
{d:n}={2:1}\\
2D Ising & {\cite{Kagawa32}} & Theory  & 0.125 & 1.75&15.0\\

\end{tabular}
\end{ruledtabular}
\end{table}

The critical exponents determined from the magnetization, along with
previously reported value and those for different known universality
classes are listed in Table I. The comparison shows that obtained
critical exponents for SIO do not fall into any known universality
class. The universality classification is generally based on the
dimensionality of lattice (d) and spin (n) system. Moreover, a two
dimensional (d = 2) nature of spin interaction is expected for
layered SIO system, which has been experimentally verified recently
{\cite{Fujiyama9, Miyazaki30}}. However, obtained $\beta$ value is
much higher than any {\it d} = 2 models {\cite{Nielsen31, Kagawa32}}
as well as previously reported $\beta$ values from muon spin
spectroscopy ($\mu$SR) (0.20) {\cite{Miyazaki30}}, neutron
diffraction (ND) (0.18) {\cite{Ye8, Chetan33}} and XRMS (0.195)
{\cite{Vale10}} techniques. Moreover, the obtained $\beta$ value is
unexpectedly large and can be linked with the presence of Griffiths
phase (not pure PM phase) which exists up to much higher
temperature, T$_G$ = 279.0(5) K than magnetic ordering temperature,
T$_C$ = 221.5 K. The similar unusual critical exponents have been
earlier reported for Griffiths ferromagnets {\cite{Jiang15, Phan16,
Triki17}} via conventional critical phenomena techniques, due to the
nonanalytic nature of Griffiths phase.

\begin{figure}
\begin{center}
\includegraphics[scale=0.30]{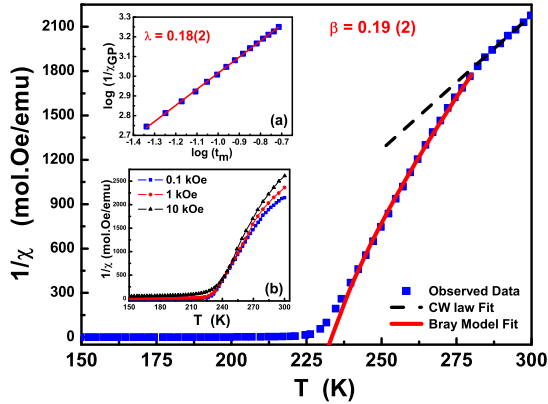}
\vspace{-1.5em}
\caption{The temperature dependence of 1/ $\chi$ at
100 Oe. The solid line is Bray model fit [Eq. 8] to Griffiths phase,
while the dashed line is Curie-Weiss law fit for reference. The
inset (a) shows 1/ $\chi$$_{GP}$(T) versus t$_m$ = (T - T$_C^R$) /
T$_C^R$ plot on log scale with power law [Eq. 5] fit. The inset (b)
shows 1/ $\chi$ (T) behavior with increasing magnetic fields from
100 Oe to 10 kOe.}
\vspace{-2.0em}
\end{center}
\end{figure}

\subsection{Griffith phase analysis}

As discussed in section III (A), negative downturn in 1/$\chi$ (T)
plot much above T$_C$ and higher experimental $\mu$$_{eff}$ value
confirm the presence of Griffiths singularity in SIO. The low field
magnetic susceptibility in the GP region, T$_C$ $<$ T $\leq$ T$_G$
follows the power law behavior {\cite{Neto34}},

\vspace{-1.5em}
\begin{equation}
\chi_{GP}^{-1} (T) \propto (T-T_C^R)^{1-\lambda},~~~~~~ 0 < \lambda < 1
\end{equation}

where T$_C^R$ is the critical temperature at which susceptibility
tends to diverge and the exponent $\lambda$ signifies the strength
of disorder. The magnetic susceptibility at 100 Oe was analyzed in
the framework of Eq. 5. The best fit gives the value of $\lambda$
and T$_C^R$ as 0.18(2) and 234.0(7) K, respectively. The fit to
the data,  plotted as 1/ $\chi$$_{GP}$ (T) versus t$_m$ = (T -
T$_C^R$) / T$_C^R$, is shown in the inset (a) of Fig. 8. The
non-zero less-than-unity exponent value further confirms the Griffiths
singularity in SIO.
The inset (b) of Fig. 8 shows the 1/$\chi$ (T) with increasing
magnetic field, which provides further evidence for GP. One can
clearly observe that negative downturn decreases with increasing
magnetic field. The applied magnetic field favors the uniform
magnetization and counteracts the disordered Griffiths phase. The
suppression of disorder can also be seen from the trend of Griffiths
exponent $\lambda$, which decreases from 0.18 (at 100 Oe) to 0.05 (at 10 kOe).
The observed behavior is consistent with earlier reports on
Griffiths ferromagnets {\cite{Nair35, Jiang36}}.

We have further analyzed GP using Bray model {\cite{Bray13,
Bray14}}, which considers the distribution of cluster sizes
characterized by a susceptibility matrix in Griffiths ferromagnet.
The inverse-susceptibility matrix eigenvalues, $\eta$/C follow the
distribution,

\begin{equation}
p(\eta)\propto \eta^{-x} exp [-A(T)/\eta]
\end{equation}

where each $\eta$ characterizes effective Curie-Weiss temperature
difference, T- $\theta$$_{eff}$ (T) for different cluster sizes. The
algebraic negative pre-factor was ignored in the original paper.
However, later on, the eigenvalues distribution [Eq. 6] was
confirmed with exponent value, x = 0.5 for diluted Ising ferromagnet
{\cite{Bray37}}. Moreover, the function A (T) vanishes at T =
T$_C^R$ and can be written as {\cite{Salamon12, Bray37}},

\begin{equation}
A(T) = A_0 \left(\frac {T}{T_C^R} - 1\right)^{2-2\beta}
\end{equation}

where A$_0$ is a constant and $\beta$ is usual critical exponent for
the system at its pure transition. As $\eta$ = T for free spins, the
average susceptibility can be written as,

\begin{equation}
\chi(T) = \frac {C}{<\eta>}= C  \frac
{\int_{0}^{T}\eta^{-1}p(\eta)d\eta}{\int_{0}^{T} p(\eta)d\eta}
\end{equation}

The fitting was performed by fixing the T$_C^R$ value to 234.0 K
obtained from power analysis Eq. (5). The best fit, shown by the
solid line in Fig 8, gives $\beta$, x and A$_0$ values as 0.19 (2),
0.52 (2) and 299 (5) K, respectively. Surprisingly, the obtained
$\beta$ (= 0.19) value is much smaller compared to MAP analysis
($\beta$ = 0.77). However, this value is in close agreement with
reported values from $\mu$SR {\cite{Miyazaki30}}, ND {\cite{Ye8, Chetan33}} and
 XRMS {\cite{Vale10}} measurements [See Table I]. The obtained
$\beta$ (= 0.19) value along with Griffith phase confirms the 2D
\emph{XYh$_4$} universality class {\cite{Taroni38}} with strong
in-plane anisotropy, which shifts the $\beta$ value from 2D XY
($\beta$ = 0.23) towards 2D Ising ($\beta$ = 0.125) interactions.
Moreover, the consistency of $\beta$ value from GP phase analysis
with previous studies suggests Bray model as a possible tool to
investigate the critical behavior for Griffiths ferromagnets.

Now, we will discuss possible origin of Griffith phase in SIO.
Two main prerequisites for the realization of Griffiths ferromagnet
are large magnetic anisotropy and intrinsic disorder.
The canted antiferromagnet nature of SIO can result in large anisotropy
between basal plane (easy axis) and c-axis (hard axis) {\cite{Hong39}}.
While in-plane strong magnetic anisotropy below T$_C$ has been observed via
Raman scattering and Torque magnetometery measurements {\cite{Gim40, Fruchter41}},
its existence well above T$_C$ is established by Vale et al. {\cite{Vale10}}.
Our critical exponent ($\beta$ = 0.19 corresponding to 2D {\it XYh$_4$} universality class)
further supports such scenario of large magnetic anisotropy in SIO.
The latter criterion for GP can be fulfilled by a number of intrinsic
sources for local disorder in SIO. As shown in the right panel of
Fig. 2, the alternate IrO$_2$ planes rotate in clockwise and anticlockwise
direction, respectively, forming a perfectly ordered structure.
However, in real samples, the sequence of rotation in alternate
layers is partially disordered as well as the rotation increases
from 11.36$^0$ (at room temperature) to 11.72$^0$ (at 10 K)
{\cite{Huang7}}. Depending on the amount of rotation, large
variation in bending modes associated with the Ir-O-Ir bond
angle can be observed {\cite{Chikara20}}, which leads to local
distribution of magnetic exchange interaction and the formation
of finite size clusters with short range magnetic order.
Moreover, the neutron diffraction measurement has shown the
presence of two crystallographically twinned magnetic domains,
which can inherently lead to formation of disordered
magnetic clusters {\cite{Chetan33}}. Apart from these, it
has been shown that two competing ordered phases can lead to
enhancement of Griffiths-like effects {\cite{Magen42, Burgy43}}.
Therefore, the competition between intralayer FM
(in basal plane) and interlayer AFM (along c-axis) interactions
in SIO can also give rise to Griffiths phase in SIO.

Next, we discuss a possible bearing of GP on the widely debated
insulating phase in SIO. SIO has found several alternative
descriptions: a Mott-Hubbard insulator {\cite{Kim6}}, in which
(on-site) Coulomb and exchange interactions are responsible for the
gap formation, or a Mott insulator {\cite{Kim1, Jackeli44}}, in
which Coulomb interaction alone leads to gap formation, or a Slater
insulator {\cite{Arita45, Li46}}, in which magnetic ordering causes
an insulating gap. However, none of these descriptions individually
provide a satisfactory explanation for the insulating nature and
metal-insulator transition (MIT) in SIO. Recently, SIO has been
termed as a ``moderately correlated" insulator in which both Mott-
and Slater-type behaviors {\cite{Hsieh47, Watanabe48}} coexist.
Although the insulating gap closes at T$_C$, indicating Slater-type
behavior, as observed by scanning tunneling spectroscopy, a
pseudo-gap still exists at higher temperatures {\cite{Li46}}.
Moreover, the optical conductivity {\cite{Moon49}} and transport
measurements {\cite{Ge50}} reveal the presence of an energy gap and
bad metallic behavior in the paramagnetic phase. All these
observations suggest a substantial contribution of the Mott-type
correlation effects, which could have a possible origin in the GP
observed here. Even though the long-range magnetic ordering vanishes
above T$_C$, short-range magnetic interactions still persist in the
GP, which could result in the formation of pseudo-gap, bad
metallicity and continuous nature of the MIT, instead of a sharp
transition in a Slater insulator. In addition, the dynamical
mean-field theory calculations have shown that the pseudo-gap can be
induced by short-range spin correlations in a Mott insulator
{\cite{Kyung51}}. Another possible origin of insulating state in SIO
is that the gap begins to form in microscopically phase-separated
regions above a bulk MIT temperature (close to T$_C$), due to the
presence of disorder in GP. Thus, a Mott-Anderson-Griffiths type MIT
may be expected in SIO similar to the proposal by Biswas et al.
{\cite{Biswas52}}; according to which SIO can be treated as an
inhomogeneous distribution of Mott insulating and Fermi liquid metal
islands, forming an electronic GP. While disorder-driven Anderson
localization have been previously suggested in SIO {\cite{Li46}} and
similar layered compound, Ba$_2$IrO$_4$ {\cite{BIO53}}, a complete
determination of its contribution to the insulating phase would
require a systematic inclusion of the disorder in SIO system.

Our study provides first experimental evidence for Griffith phase in
SIO within the temperature range of T$_C$ (=  221.5 K) $<$ T $\leq$
T$_G$ (= 279 K). The manifestation of Griffith ferromagnetic phase
have been observed in few recent experiments. For example, the XRMS
measurements have shown that the magnetic correlations exist in
paramagnetic phase above T$_C$ {\cite{Fujiyama9, Vale10}}. Moreover,
a recent Raman study has reported the presence of two-magnon
scattering well above T$_C$ {\cite{Gretarsson54}}. This
provides the evidence for strong pseudospin excitations in
paramagnetic state of SIO and the quenching of such fluctuations
lead to sharp, dispersive magnon exciton modes below T$_C$. Still,
further understanding of such novel magnetic phase in iridates
demands more experimental and theoretical studies. The experimental
techniques with high magnetic sensitivity (like electron spin
resonance) are necessary to probe SIO near critical region. On the
theoretical front, the models incorporating local disorder (like
octahedral rotation), magnetic anisotropy and FM/AFM phase
competition are needed to understand the GP and the insulating
nature of SIO.

\vspace{-1.5em}
\section{Conclusion}

In conclusion, the existence of  Griffiths phase (GP) and its
influence on critical phenomena in layered Sr$_2$IrO$_4$
ferromagnet are reported in this work.
The inverse susceptibility, 1/$\chi$(T) shows downward deviation
from CW behavior much above T$_C$, showing the existence of GP
in SIO. The GP was confirmed from the power-law behavior of 1/$\chi$ (T)
with less-than-unity exponent value, $\lambda$ = 0.18(2). The
detailed isotherm analysis around T$_C$ using modified Arrott plot
reveals second order phase transition with critical parameters,
$\beta$ = 0.77(1), $\gamma$ = 1.59(2) and $\delta$ = 3.06 (4) with T$_C$ = 221.5(1) K,
in consistent with magneto-caloric study. However, the obtained $\beta$
value is unrealistically large and has been associated with
{\it ferromagnetic-Griffiths} phase transition.
Further analysis of GP using Bray model gives reliable $\beta$
[0.19(2)] value for Sr$_2$IrO$_4$. The obtained $\beta$ value
corresponding to 2D \emph{XYh$_4$} universality class
confirms the strong in-plane anisotropy, supporting
presence of GP in the system.  This study proposes Bray model as
a possible tool to investigate the critical behavior for Griffiths
ferromagnets via dc magnetization study.
Furthermore, the study suggests that the GP plays an important role in
Mott-type short-range correlations above T$_C$ and disorder driven Anderson
localization effects, which may be the possible origin of unconventional
insulating nature of Sr$_2$IrO$_4$.

\vspace{-1.5em}
\section{ACKNOWLEDGMENT}

A. Rathi thanks University Grants Commission (UGC), India for
providing the junior research fellowship (JRF) to carry out this
research work and S. Perween thanks Department of Science and
technology (DST), India for providing INSPIRE fellowship.

\end{document}